\documentclass[smallcondensed,final]{svjour3}
\usepackage[numbers,sort&compress]{natbib}
\usepackage{graphicx}
\usepackage{amsmath}
\usepackage{amssymb}
\usepackage{hyperref}

\begin{document}

\title{{Hydrodynamic Fluctuations in Laminar Fluid Flow. II. Fluctuating Squire Equation}}
\titlerunning{Hydrodynamic Fluctuations in Laminar Fluid Flow. II.}
\author{J. M. Ortiz de Z\'{a}rate \and J. V. Sengers}

\institute{J. M. Ortiz de Z\'{a}rate \at {Departamento de F\'{\i}sica Aplicada I, Facultad de F\'{\i}sica,}\\ {Universidad Complutense, 28040 Madrid, Spain.}\\ \email{jmortizz@fis.ucm.es}
    \and J. V. Sengers \at {Institute for Physical Science and Technology}\\{and Burgers Program for Fluid Dynamics,}\\{University of Maryland. College Park, Maryland 20742-8510, USA.}\\ \email{sengers@umd.edu}}

\maketitle

\begin{abstract}
We use fluctuating hydrodynamics to evaluate the enhancement of thermally excited fluctuations in laminar fluid flow using plane Couette flow as a representative example. In a previous publication [J. Stat. Phys. {\bf144} (2011) 774] we derived the energy amplification arising from thermally excited wall-normal fluctuations by solving a fluctuating Orr-Sommerfeld equation. In the present paper we derive the energy amplification arising from wall-normal vorticity fluctuation by solving a fluctuating Squire equation. The thermally excited wall-normal vorticity fluctuations turn out to yield the dominant contribution to the energy amplification. In addition, we show that thermally excited streaks, even in the absence of any externally imposed perturbations, are present in laminar fluid flow.
\end{abstract}


\section{Introduction}

The presence of gradients, such as temperature gradients, concentration gradients, or velocity gradients, always causes non-equilibrium enhancements of thermal fluctuations that are spatially long ranged, even when the system is far away from any hydrodynamic instability~\cite{DorfmanKirkpatrickSengers,BOOK}. The present paper is part of a detailed study of the nature of thermally excited fluctuations in laminar fluid flow using plane Couette flow as a representative example. It has been verified that fluctuating hydrodynamics, originally developed for thermal fluctuations in equilibrium states~\cite{LandauLifshitz,FoxUhlenbeck1}, can be extended to deal with thermal fluctuations in non-equilibrium states~\cite{BOOK}. In fluctuating hydrodynamics the usual deterministic hydrodynamic equations are supplemented with random dissipative fluxes of thermal (natural) origin. In the case of laminar fluid flows one has to consider a random stress tensor to account for intrinsic thermal noise that always will be present. This noise will be amplified by the presence of a velocity gradient. Energy amplification induced by the flow has attracted the attention of many investigators. In many of the studies reported in the literature the noise is not of thermal origin and does not obey a fluctuation-dissipation relation~\cite{FarrellIoannou,FarrellIoannou2,BamiehDahleh,JovanovicBamieh,EckhardtPandit}. Fluctuating hydrodynamics provides a systematic method for assessing the nature of spontaneous fluctuations in laminar flow induced by intrinsic noise. The application of fluctuating hydrodynamics to shear flows has been initiated by some previous investigators, but without considering confinement effects~\cite{TremblayEtAl,LD85,LDD89,LD02,WS03}. However, the long-ranged nature of the fluctuations is highly anisotropic and for certain directions of the wave vector the fluctuations encompass the entire fluid system, so that boundary effects will affect these fluctuations.

Previously we have derived the appropriate fluctuating hydrodynamics equations for laminar fluid flow. Specifically, we have shown how the thermally excited wall-normal velocity fluctuations can be described by a stochastic Orr-Sommerfeld equation~\cite{miCouette} and the thermally excited vorticity fluctuations by a stochastic Squire equation~\cite{miCouette2}. Accounting for realistic boundary conditions we obtained solutions of the stochastic Sommerfeld and Squire equations based on semi-quantitative Galerkin approximations\cite{miCouette,miCouette2,miJNNFM}. We now have derived more exact solutions of these stochastic equations in terms of an expansion of the eigenfunctions (hydrodynamic modes) of the hydrodynamic operator. The more exact solution of the stochastic Orr-Sommerfeld equation for the wall-normal velocity fluctuations has been presented in a previous article in this series, to be referred to as paper~I~\cite{miORR}. We found that the flow-induced enhancement of the wall-normal velocity fluctuations and the resulting energy amplification increases with the Reynolds number Re approximately as Re$^2$. The actual enhancement of the velocity fluctuations strongly depends on the wave number. For large wave numbers (in the bulk of the fluid), this enhancement varies as the fourth power of the inverse wave number, independent of any boundary conditions. For small wave numbers the enhancement vanishes as the square of the wave number due to the presence of boundaries. Our previous approximate solution based on a Galerkin approximation~\cite{miCouette} did reproduce the correct dependence of the non-equilibrium enhancement of the fluctuations on the wave number but underestimated the magnitude of the enhancement at intermediate wave numbers~\cite{miORR}.

The present paper is concerned with an analysis of the solution of the stochastic Squire equation for the wall-normal vorticity fluctuations. We shall proceed as follows. In Sect.~\ref{S02} we recall the expressions for the stochastic Orr-Sommerfeld and Squire equations in terms of suitable dimensionless variables. The Squire equation for the wall-normal vorticity fluctuations includes a coupling with the Orr-Sommerfeld equation for the wall-normal velocity fluctuations. Hence, the solution of the stochastic Squire equation for the wall-normal vorticity fluctuations to be obtained in the present paper will depend on the solution of the stochastic Orr-Sommerfeld equation for the wall-normal velocity fluctuations obtained in our preceding paper~\cite{miORR}. In Sect.~\ref{S02B} we describe the procedure for solving the stochastic Squire equation. For this purpose we expand the solution in terms of the eigenfunctions of the linear hydrodynamic operator associated with the Squire equation. In Sect.~\ref{S3} we derive the corresponding hydrodynamic modes and decay rates. In Sect.~\ref{S04} we then obtain the expressions for both the equilibrium and nonequilibrium contributions to the intensity of the vorticity fluctuations, while in Sect.~\ref{S05} we deduce the nonequilibrium energy amplification arising from these vorticity fluctuations. The nonequilibrium energy enhancement turns out to be proportional to the square of the Reynolds number. Specifically, we evaluate the nonequilibrium energy amplification associated with the vorticity fluctuations for fluctuations with wave vector in the spanwise direction which appears to be the most interesting case. We also find good agreement between the exact solution, obtained in this paper, and the semi-quantitative solution previously obtained in a Galerkin approximation~\cite{miCouette2}. We conclude with some general comments in Sect.~\ref{S07}.

\section{Fluctuating Hydrodynamics of Shear Flow. The Stochastic Orr-Sommerfeld and Squire Equations\label{S02}}

We consider a liquid with uniform temperature $T$ under incompressible laminar flow  with uniform density $\rho$ between two horizontal plates separated by a distance $2L$. As in our previous publication~\cite{miORR} we adopt a coordinate system with the $X$-axis in the streamwise direction, the $Y$-axis in the spanwise direction, and the $Z$-axis in the wall-normal direction~\cite{DrazinReid}. Thus the mean flow velocity  $\mathbf{v}_0=\{\dot{\gamma} z,0,0\}$ is in the $X$-direction with $\dot{\gamma}$ representing a constant shear rate in the $Z$-direction. The bounding upper plate, at the position $z=+L$, moves in the positive $X$-direction with velocity $\dot{\gamma}L$, while the lower plate, at the position $z=-L$, moves with the same velocity in the opposite direction. This flow configuration is commonly referred to as plane Couette flow. It is convenient to use a dimensionless position variable $\mathbf{r}$, measured in terms of the length $L$, a dimensionless time $t$ obtained by multiplying the actual time with the shear rate $\dot{\gamma}$, a dimensionless fluid velocity $\mathbf{v}$ in terms of the product $\dot{\gamma}L$, and a dimensionless stress tensor $\boldsymbol{\Pi}$ in terms of $\rho{L}^2\dot{\gamma}$.

We want to study velocity fluctuations around the stationary flow solution of the Navier-Stokes equation. Specifically, we are interested in fluctuations of thermal origin, \emph{i.e.}, fluctuations resulting from the intrinsically stochastic nature of molecular motions. Such fluctuations are always present and are unavoidably linked to any dissipative processes that are present in the system, Newton's law of viscosity in the present case. Fluctuating hydrodynamics provides a general and systematic framework for describing such thermal fluctuations, even when the system is in a non-equilibrium state~\cite{DorfmanKirkpatrickSengers,BOOK}. The idea is that the linear phenomenological laws representing dissipation in the system are to be supplemented with random dissipative fluxes (thermal noise), whose statistical properties are given by the fluctuation-dissipation theorem, see \emph{e.g.}~\cite{Kubo}. The goal is then to obtain the correlation functions of the fluctuating thermodynamic fields, velocity fluctuations $\delta\mathbf{v}$ in our case, in terms of the statistical properties of the thermal noise, the stochastic stress tensor $\delta\boldsymbol{\Pi}$ in our case. Implementing this procedure we have shown in previous publications that the fluctuations $\delta{v}_z$ of the wall-normal velocity component $v_z$ satisfy a stochastic Orr-Sommerfeld equation~\cite{miCouette}
\begin{equation}
\partial_{t}(\nabla^2\delta{v}_z)+
z~\partial _{x}(\nabla^2\delta{v}_z) - \frac{1}{\mathrm{Re}}
\nabla^{4}(\delta{v}_z) = -
\left\{\boldsymbol{\nabla}\times\boldsymbol{\nabla}\times\left[\boldsymbol{\nabla}\left(\delta\boldsymbol\Pi
\right)\right]\right\}_z,\label{Eq09A}
\end{equation}
and the vorticity fluctuations $\delta{w}_z=\partial_x\delta{v}_y-\partial_y\delta{v}_x$  a stochastic Squire equation~\cite{miCouette2}
\begin{equation}
\partial_{t}(\delta{w}_z)+
z~\partial _{x}(\delta{w}_z) -
\partial_y\delta{v}_z - \frac{1}{\mathrm{Re}}
\nabla^{2}(\delta{w}_z) =
\left\{\boldsymbol{\nabla}\times\left[\boldsymbol{\nabla}\left(\delta\boldsymbol\Pi
\right)\right]\right\}_z. \label{Eq09B}
\end{equation}
In these equations Re is the Reynolds number
\begin{equation}
\text{Re}=\frac{\rho\dot{\gamma} L^2}{\eta}
\end{equation}
with $\eta$ being the shear viscosity of the fluid. By combining Eqs.~\eqref{Eq09A} and~\eqref{Eq09B} with the incompressibility assumption, $\boldsymbol{\nabla}\cdot\delta\mathbf{v}$, one can obtain the three components of the fluctuating velocity field $\delta\mathbf{v}$.

The difference between the stochastic Orr-Sommerfeld equation~\eqref{Eq09A} and the stochastic Squire equation~\eqref{Eq09B} and their deterministic counterparts~\cite{DrazinReid,SchmidHenningson} is the presence of noise terms on the right-hand side (RHS). These additive noise terms appear as derivatives of the random stress $\delta\boldsymbol{\Pi}$; their correlation functions can be deduced from the fluctuation dissipation theorem which in this case reads~\cite{BOOK}
\begin{equation}
\langle \delta \Pi _{ij}(\mathbf{r},t)\cdot \delta \Pi _{kl}(\mathbf{r}%
^{\prime },t^{\prime })\rangle =2\tilde{S}  \left(
\delta_{ik}\delta _{jl}+\delta _{il}\delta _{jk}\right) \delta (\mathbf{r}-\mathbf{%
r}^{\prime })~\delta (t-t^{\prime }).\label{FDT1}
\end{equation}
In this equation $\tilde{S}$ is the dimensionless strength of the thermal noise~\cite{miORR}:
\begin{equation}\label{Eq04B}
\tilde{S}=\frac{k_\text{B} T}{\dot{\gamma}^3 L^7}
\frac{\eta}{\rho_0^2} = \frac{k_\mathrm{B} T}{\rho_0 L^3}
\frac{1}{\dot{\gamma}^2 L^2} \frac{1}{\mathrm{Re}}.
\end{equation}
where $k_\text{B}$ is Boltzmann's constant. We note that the actual correlation function for the fluctuating stress tensor only depends on the properties of the fluid, namely temperature, density, and viscosity; the shear rate and the Reynolds number only appears in~\eqref{Eq04B} as a consequence of the manner in which the stress has been made dimensionless.

Equations~\eqref{Eq09A} and~\eqref{Eq09B} form a pair of coupled stochastic differential equations which have to be solved for the velocity and vorticity fluctuations subject to appropriate no-slip boundary conditions:
\begin{equation}\label{BC1}
\delta{v}_z(\mathbf{r},t)=\partial_z\delta{v}_z(\mathbf{r},t)=\delta{w}_z(\mathbf{r},t)=0,\hspace*{20pt}\text{at}~z=\pm{1}.
\end{equation}

In our earlier work we have accounted for these boundary conditions by solving Eqs.~\eqref{Eq09A} and~\eqref{Eq09B} using a Galerkin method that allowed for analytical but approximate expressions for the correlation functions of the wall-normal velocity and vorticity fluctuations~\cite{miCouette,miCouette2}. We have summarized these approximate solutions of the two stochastic equations in a subsequently publication~\cite{miJNNFM}. While the Galerkin-approximation technique enabled us to obtain relatively simple analytical results, the approximation is somewhat uncontrolled. Hence, we found it desirable to compare the approximate analytical solutions with exact numerical solutions that can be obtained through expansions in eigenfunctions of the hydrodynamic operators. We have implemented this project for the solution of the Orr-Sommerfeld stochastic equation~\eqref{Eq09A}, yielding the correlation function for the wall-normal velocity fluctuations in paper~I in this series~\cite{miORR}. In the present paper we analyze the solution of the stochastic Squire equation~\eqref{Eq09B} for the wall-normal vorticity fluctuations. There is an important difference between the Orr-Sommerfeld equation~\eqref{Eq09A} and the Squire equation~\eqref{Eq09B}. The solution of the Orr-Sommerfeld equation for the velocity fluctuations is independent of the solution of the Squire equation for the vorticity fluctuations. On the other hand, the Squire equation~\eqref{Eq09B} is coupled with the Orr-Sommerfeld equation though the presence of the term $\partial_y\delta{v}_z$ in Eq.~\eqref{Eq09B}. Physically it means that the wall-normal velocity fluctuations involve only a coupling between the same (viscous) hydrodynamic mode (at different wave numbers) to which we have referred as "self-coupling"~\cite{miCouette}. However, the solution of the Squire equation for the wall-normal vorticity fluctuations is related to two mode-coupling mechanisms: a self-coupling between vorticity fluctuations and a cross-coupling between velocity and vorticity fluctuations. We shall recover our earlier observation~\cite{miCouette2} that the self-coupling mechanism in the Squire equation determines the intensity of the vorticity fluctuations in equilibrium and that the cross-coupling mechanism in the Squire equation determines the nonequilibrium intensity enhancement of the wall-normal vorticity fluctuations.

To put our present work in the context of previous investigations of the effect of externally imposed stochastic forcing on shear flows we may mention the following. In their original papers, Farrell and Ionannou~\cite{FarrellIoannou,FarrellIoannou2} and Bamieh and Dahleh~\cite{BamiehDahleh} introduced a stochastic forcing term directly into the RHS of the Orr-Sommerfeld and Squire equations~\eqref{Eq09A} and~\eqref{Eq09B} instead of the thermal forcing incorporated by us. More recently, Jovanovic and Bamieh~\cite{JovanovicBamieh} introduced the forcing in the more basic Navier-Stokes equation, resulting in a more transparent interpretation of the random terms as actual forces. Moreover, Jovanovic and Bamieh~\cite{JovanovicBamieh} allowed for some flexibility in the spatial spectrum of the noise, distinguishing between structured and unstructured noise. However, in spite of this flexibility, thermal noise was not included as a particular case considered by the authors~\cite{JovanovicBamieh}. Indeed, from the flutuation-dissipation relation~\eqref{FDT1}, it follows that the spatial spectrum of the resulting stochastic forcing in the RHS of Eqs.~\eqref{Eq09A} and~\eqref{Eq09B} is not contained in the expressions of Jovanovic and Bamieh~\cite{JovanovicBamieh}. Hence, the results obtained here will differ from previous investigations.

In terms of the input-output nomenclature that, borrowed from the dynamics and control literature, has become popular lately~\cite{JovanovicBamieh,HwangCossu1}; we take as input the thermal noise and our output (through the linearized Navier-Stokes equations) is the wall-normal vorticity autocorrelation function.

\section{Procedure for Solving the Stochastic Squire Equation\label{S02B}}

Just as the procedure used for solving the Sommerfeld-Orr equation~\cite{miORR}, to solve the Squire equation~\eqref{Eq09B} we apply a Fourier transform in time and in the horizontal $XY$-plane parallel to the walls:
\begin{equation}
\left[\mathrm{i}\omega+\mathcal{H}\right]\cdot\delta{w}_z(\omega, \mathbf{q}_\parallel,z)=
\mathrm{i}q_y\delta{v}_z(\omega,\mathbf{q}_\parallel,z)+S_z(\omega,\mathbf{q}_\parallel,z),\label{Eq10}
\end{equation}
where $\omega$ is the frequency of the fluctuations $\delta{w}_z$ and $\mathbf{q}_\parallel=\{q_x,q_y\}$ the corresponding wave vector in the plane parallel to the walls. In Eq.~\eqref{Eq10}, $\mathcal{H}$ represents a linear hydrodynamic operator:
\begin{equation}\label{E49}
\mathcal{H}=\mathrm{i} zq_x - \dfrac{1}{\mathrm{Re}}\left(\partial_z^2-q_\parallel^2\right),
\end{equation}
The RHS of Eq.~\eqref{E49} contains two stochastic forcing terms. The first one, $\text{i}q_y\delta{v}_z$, accounts for a stochastic forcing originating from the wall-normal velocity fluctuations, which can be represented in terms of the exact solution of the Orr-Sommerfeld equation with thermal forcing derived in paper~I~\cite{miORR}. The second forcing term $S_z$ is given by the Fourier transform, in time and in the $XY$-plane, of the combination of derivatives of the random stress $\delta\boldsymbol{\Pi}$ in the RHS of Eq.~\eqref{Eq09B}:
\begin{equation}\label{E6}
S_z=\mathrm{i}\partial_z[q_x\delta\Pi_{zy}-q_y\delta\Pi_{zx}]-q_x^2\delta\Pi_{xy}+q_y^2\delta\Pi_{yx}+q_xq_y[\delta\Pi_{xx}-\delta\Pi_{yy}].
\end{equation}
Since the hydrodynamic operator is linear we find the solution of Eq.~\eqref{Eq10} by expanding in a set of right eigenfunctions, $R_N(\mathbf{q}_\parallel,z)$, or hydrodynamic modes as:
\begin{equation}
\delta{\omega}_z(\omega,\mathbf{q}_\parallel,z)= \sum_{N=0}^\infty G_N(\omega,\mathbf{q}_\parallel)
~R_N(\mathbf{q}_\parallel,z),\label{EX33}
\end{equation}
where the hydrodynamic modes are the solution of:
\begin{equation}
\mathcal{H}\cdot{R}_N(\mathbf{q}_\parallel,z) =
\Gamma_N(\mathbf{q}_\parallel)~{R}_N(\mathbf{q}_\parallel,z)\label{E50}
\end{equation}
with $\Gamma_N(\mathbf{q}_\parallel)$ being the corresponding eigenvalue or decay rate. In Eq.~\eqref{E50} $R_N(\mathbf{q}_\parallel,z)$ must satisfy the boundary conditions $R_N(\mathbf{q}_\parallel,\pm1)=0$. In Eqs.~\eqref{EX33}-\eqref{E50} we anticipated the fact, to be discussed in more detail in Sect.~\ref{S3}, that the right eigenvalue problem of the Squire operator, Eq.~\eqref{E50}, has indeed an infinite numerable set of solutions. Next, to evaluate the coefficients $G_N(\omega,\mathbf{q}_\parallel)$ of the series expansion~\eqref{EX33} we use the property that the complex conjugates ${R}_N^*(\mathbf{q}_\parallel,z)$ are the left eigenfunctions of the Squire operator~\eqref{E49}, with corresponding eigenvalues $\Gamma_N^*(\mathbf{q}_\parallel)$. Indeed, by using the boundary conditions~\eqref{BC1} it can be readily shown that the adjoint of the operator $\mathcal{H}$ is simply its complex conjugate and the left eigenfunction is just the complex conjugate of the right eigenfunction. As a consequence, the biorthogonality~\cite{CourantHilbert} condition reads:
\begin{equation}\label{E60}
\int_{-1}^{1}
{R}_M(\mathbf{q}_\parallel,z)~{R}_N(\mathbf{q}_\parallel,z)
~ dz= B_N(\mathbf{q}_\parallel)~\delta_{NM}.
\end{equation}
Equation~\eqref{E60} determines the normalization of the eigenfunctions. Next, we evaluate the coefficients $G_N(\omega,\mathbf{q}_\parallel)$ in the expression~\eqref{EX33} for $\delta{\omega}_z(\omega,\mathbf{q}_\parallel,z)$ by substituting Eq.~\eqref{EX33} into Eq.~\eqref{Eq10}. The resulting expression is then projected onto ${R}_M(\mathbf{q}_\parallel,z)$ and using the biortogonality condition~\eqref{E60}, one readily obtains:
\begin{equation}\label{E10}
G_N^{\mathrm{R}}(\omega,\mathbf{q}_\parallel)=\frac{F_N(\omega,\mathbf{q}_\parallel)}
{B_N(\omega,\mathbf{q}_\parallel)[\mathrm{i}\omega +
\Gamma_N(\mathbf{q}_\parallel)]},
\end{equation}
where $F_N(\omega,\mathbf{q}_\parallel)$ are the projections of the RHS of Eq.~\eqref{Eq10} onto the ${R}_N(\mathbf{q}_\parallel,z)$ functions, namely:
\begin{equation}\label{E11}
F_N(\omega,\mathbf{q}_\parallel) = \int_{-1}^{1}
R_N(\mathbf{q}_\parallel,z)\left[\mathrm{i}q_y\delta{v}_z(\omega,\mathbf{q}_\parallel,z)+S_z(\omega,\mathbf{q}_\parallel,z) \right]~ dz.
\end{equation}
For the evaluation of the vorticity fluctuations, we need the correlation functions $\langle{F}_N^*(\omega,\mathbf{q}_\parallel) \cdot{F}_M(\omega^\prime,\mathbf{q}_\parallel^\prime)\rangle$. These, in turn, can be deduced from the correlations:
\begin{subequations}\label{E12}
\begin{align}
\langle\delta{v}_z^*(\omega,\mathbf{q}_\parallel,z) \cdot\delta{v}_z(\omega^\prime,\mathbf{q}_\parallel^\prime,z^\prime)\rangle &= (2\pi)^3~\delta(\omega-\omega^\prime)~\delta(\mathbf{q}_\parallel-\mathbf{q}_\parallel^\prime)~C_{zz}(\omega,\mathbf{q}_\parallel,z,z^\prime),\label{E12A}\\
\langle{S}_z^*(\omega,\mathbf{q}_\parallel,z) \cdot{S}_z(\omega^\prime,\mathbf{q}_\parallel^\prime,z^\prime)\rangle &=(2\pi)^3~\delta(\omega-\omega^\prime)~\delta(\mathbf{q}_\parallel-\mathbf{q}_\parallel^\prime)\notag\\&\hspace*{60pt}\times2\tilde{S}~q_\parallel^2(q_\parallel^2+\partial_z\partial_{z^\prime})\delta(z-z^\prime), \label{E12B}\\
\langle{S}_z^*(\omega,\mathbf{q}_\parallel,z) \cdot\delta{v}_z(\omega^\prime,\mathbf{q}_\parallel^\prime,z^\prime)\rangle &=
\langle\delta{v}_z^*(\omega,\mathbf{q}_\parallel,z) \cdot{S}_z(\omega^\prime,\mathbf{q}_\parallel^\prime,z^\prime)\rangle=0.\label{E12C}
\end{align}
\end{subequations}
Equation~\eqref{E12A} represents the solution of the stochastic Orr-Sommerfeld equation obtained in paper~I~\cite{miORR} (see Eq.~(24) in~\cite{miORR}), where explicit expressions for the function $C_{zz}(\omega,\mathbf{q}_\parallel,z,z^\prime)$ in terms of the hydrodynamic modes and decay rates of the Orr-Sommerfeld operator have been presented. We shall not repeat those expression here, although they shall be used in some of the following calculations. Equation~\eqref{E12C} is a consequence of the fact that the random noise terms in the stochastic Orr-Sommerfeld and Squire equations are uncorrelated, see Eq.~(18) in Ref.~\cite{miCouette2}. Finally, Eq.~\eqref{E12B}, which is new in this paper, is readily obtained from the definition~\eqref{E6} of ${S}_z(\omega,\mathbf{q}_\parallel,z)$ and the fluctuation-dissipation theorem~\eqref{FDT1} for the random stress tensor. The expression for the prefactor $\tilde{S}$ is given by Eq.~\eqref{Eq04B}.

With the help of Eqs.~\eqref{E12}, the correlations $\langle{F}_N^*(\omega,\mathbf{q}_\parallel)\cdot{F}_M(\omega^\prime,\mathbf{q}_\parallel^\prime)\rangle$ between the coefficients defined in Eq.~\eqref{E11}, can be expressed as:
\begin{equation}\label{E13}
\langle{F}_N^*(\omega,\mathbf{q}_\parallel) \cdot{F}_M(\omega^\prime,\mathbf{q}_\parallel^\prime)\rangle= \left[\Xi^\text{(E)}_{NM}(\mathbf{q}_\parallel) + \Xi^\text{(NE)}_{NM}(\omega,\mathbf{q}_\parallel)\right]
(2\pi)^3~\delta(\omega-\omega^\prime)~\delta(\mathbf{q}_\parallel-\mathbf{q}_\parallel^\prime),
\end{equation}
with mode-coupling coefficients
\begin{subequations}\label{E14}
\begin{align}
\Xi^\text{(E)}_{NM}(\mathbf{q}_\parallel)&=2\tilde{S}~q_\parallel^2 \int_{-1}^{1}dz~R_N^*(q_\parallel,z) \left[q_\parallel^2-\partial_z^2\right] R_M(q_\parallel,z),\label{E14A}\\
\Xi^\text{(NE)}_{NM}(\omega,\mathbf{q}_\parallel)&=q_y^2\iint_{-1}^{1}\hspace*{-6pt}dz~dz^\prime~R_N^*(q_\parallel,z) ~C_{zz}(\omega,\mathbf{q}_\parallel,z,z^\prime)~R_M(q_\parallel,z^\prime).\label{E14B}
\end{align}
\end{subequations}
The first expression,~\eqref{E14A}, is obtained from Eqs.~\eqref{E11} and~\eqref{E12B} if one performs integrations by parts to move the derivatives in~\eqref{E12B} from the delta function to the hydrodynamic modes, and uses the boundary conditions  $R_N(\mathbf{q}_\parallel,\pm1)=0$. In Eqs.~\eqref{E13} and~\eqref{E14} we have introduced superscripts (E) and (NE) to distinguish between the two contributions, anticipating the fact (to be discussed at length later) that the first set of mode-coupling coefficients will contribute only to the equilibrium equal-time vorticity fluctuations, while the second set of mode-coupling coefficients (due to the coupling of wall-normal velocity and vorticity fluctuations) contains the nonequilibrium amplification (enhancement) of those fluctuations.

Equation~\eqref{E14A} for $\Xi^\text{(E)}_{NM}(\mathbf{q}_\parallel)$ can be further transformed by integrating by parts while making use of the eigenvalue problem~\eqref{E50}, to obtain an expression more useful for future use:
\begin{equation}\label{E15}
\Xi^\text{(E)}_{NM}(\mathbf{q}_\parallel)=\tilde{S}\text{Re}~q_\parallel^2\left[\Gamma_N^*(q_\parallel)+ \Gamma_M(q_\parallel)\right] \int_{-1}^{1}d\xi~R_N^*(q_\parallel,\xi)~R_M(q_\parallel,\xi),
\end{equation}
where we renamed the integration variable as $\xi$.

\section{Hydrodynamic Modes and Decay Rates\label{S3}}

The hydrodynamic modes $R_N(\mathbf{q}_\parallel,z)$ of the operator $\mathcal{H}$ are obtained by solving Eq.~\eqref{E50} with the appropriate boundary conditions, \emph{i.e.}, $R_N(\mathbf{q}_\parallel,\pm1)=0$. In view of the definition~\eqref{E49} of $\mathcal{H}$, the general solution of Eq.~\eqref{E50} can be expressed as a linear combination of Airy functions. The boundary condition at $z=1$ can be easily accommodated by a convenient selection of the coefficients. Then the solution of Eq.~\eqref{E50} satisfying the boundary condition at $z=1$ can be written in the form~\cite{Romanov}:
\begin{multline}\label{E72}
R_N(z) = \mathrm{Bi}\left[(q_x\mathrm{Re})^{1/3}(a_N-\mathrm{i})\right]~\mathrm{Ai}\left[(q_x\mathrm{Re})^{1/3}(a_N-\mathrm{i}z)
\right] \\-
\mathrm{Ai}\left[(q_x\mathrm{Re})^{1/3}(a_N-\mathrm{i})\right]~\mathrm{Bi}\left[(q_x\mathrm{Re})^{1/3}(a_N-\mathrm{i}z)
\right].
\end{multline}
In principle, the parameter $a_N$ in~\eqref{E72} can be any complex number. The decay rate of the hydrodynamic mode~\eqref{E72} is expressed in terms of this parameter $a_N$ as:
\begin{equation}
\Gamma_N(\mathbf{q}_\parallel) = q_x~a_N(\mathbf{q}_\parallel)
+ \frac{q_\parallel^2}{\mathrm{Re}}.\label{Eq51}
\end{equation}
For the hydrodynamic mode~\eqref{E72} to satisfy the boundary condition at $z=-1$ we need to impose the  condition:
\begin{multline}\label{E74}
0=\mathrm{Bi}\left[(q_x\mathrm{Re})^{1/3}(a_N-\mathrm{i})\right]
~\mathrm{Ai}\left[(q_x\mathrm{Re})^{1/3}(a_N+\mathrm{i})\right]
\\-\mathrm{Ai}\left[(q_x\mathrm{Re})^{1/3}(a_N-\mathrm{i})\right]
~\mathrm{Bi}\left[(q_x\mathrm{Re})^{1/3}(a_N+\mathrm{i})\right].
\end{multline}
Because of the oscillatory character of the Airy functions, Eq.~\eqref{E74} has an infinite numerable set of complex roots $a_N$. This fact has been anticipated in Eq.~\eqref{EX33}, and the index $N$ has been used throughout to distinguish among the various modes. In principle, the decay rates $a_N$ depend on the parallel wave vector $\mathbf{q}_\parallel=\{q_x,q_y\}$ and on the Reynolds number Re. However, as a consequence of the structure of Eq.~\eqref{E74}, the decay rates only depend on the magnitude $q_\parallel$ of the wave vector $\mathbf{q}_\parallel$ in the plane parallel to the plates and on an effective Reynolds number
\begin{equation}\label{EN21}
\overline{\mathrm{Re}}=\mathrm{Re}\cos\varphi,
\end{equation}
where $\varphi$ is the azimuthal angle of the wave vector $\mathbf{q}_\parallel$, measured with respect to the stream-wise $X$-direction. Hence, $q_x\text{Re}=q_\parallel\overline{\text{Re}}$. This simplification is commonly referred to as Squire symmetry~\cite{DrazinReid}.  As was discussed in detail in the preceding paper~\cite{miORR}, the eigenvalues and the eigenfunctions of the Orr-Sommerfeld hydrodynamic operator obey the same Squire symmetry. However, in contrast to the decay rates of the Orr-Sommerfeld hydrodynamic operator, the decay rates $a_N$ of the Squire hydrodynamic operator have an additional symmetry property, namely
\begin{equation}\label{EN22}
a_N(q_\parallel,\overline{\mathrm{Re}})=a_N(q_\parallel/\lambda,\overline{\mathrm{Re}}\lambda)
\end{equation}
for any real parameter value $\lambda$. As a consequence, it is sufficient to determine the solutions of Eq.~\eqref{E74} for a single value of the effective Reynolds number, such as $\overline{\mathrm{Re}}=1$.
\begin{figure}
\begin{center}
\resizebox{0.60\textwidth}{!}{\includegraphics{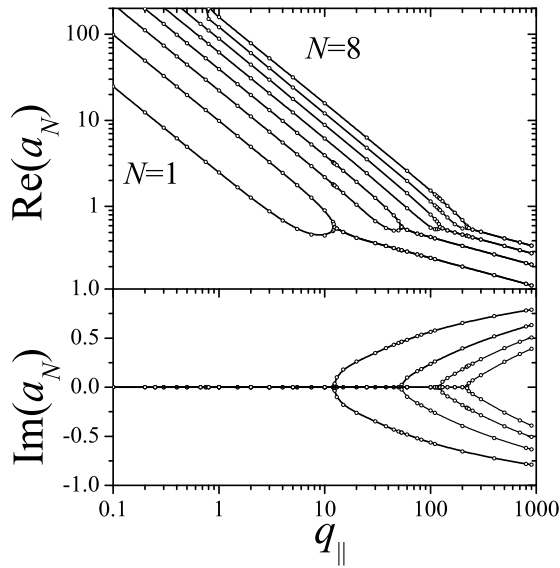}}
\end{center}
\caption{Real (upper panel) and imaginary (lower panel) parts of the eight slowest decay rates as a function of $q_\parallel$, for effective Reynolds number $\overline{\mathrm{Re}}=1$.}\label{F1}
\end{figure}

In Fig.~\ref{F1} we show, as a function of the wavenumber $q_\parallel$, the decay rates of the Squire hydrodynamic operator $\mathcal{H}$ for $\overline{\mathrm{Re}}=1$. Use of the explicit exact expression~\eqref{E72} for the eigenfunctions (and, hence, Eq.~\eqref{E74}) enables us for a much simpler computation of the data displayed in Fig.~\ref{F1} when compared to a direct numerical integration of Eq.~\eqref{E50} performed by Gustavsson and Hultgren~\cite{GustavssonHultgren}. As expected, the results are the same: see, \emph{e.g.}, Fig.~1 of Ref.~\cite{GustavssonHultgren}. For small $q_\parallel$ the decay rates are real numbers and a simple perturbative calculation allows to obtain the first terms in a series expansion in powers of $q_\parallel$, namely:
\begin{equation}
\left.a_N(q_\parallel)\right|_{\overline{\text{Re}}=1} = \frac{N^2\pi^2}{4~q_\parallel} + \mathcal{O}(q_\parallel),\hspace*{50pt}\text{for}~N=1,2,3\dots
\end{equation}
For larger $q_\parallel$ and depending on the order $N$, the decay rates merge in pairs of complex conjugate numbers. For even larger $q_\parallel\to\infty$ the real part of $a_N$ decay to zero as $q_\parallel^{-1/3}$, while the imaginary parts of each pair approach $\pm\mathrm{i}$. As explained above, this general landscape of decay rates is maintained for any Reynolds number because of the scaling relation~\eqref{EN22}. In particular, none of the decay rates becomes zero for any value of the wave vector or the Reynolds number. The same is true for the Orr-Sommerfeld equation and, as is well known, there is no linear hydrodynamic instability in plane Couette flow.

One difference with the landscape of eigenvalues of the Orr-Sommerfeld operator is that there is no transient merging of eigenvalues in a limited ``window" of wave numbers, nor crossing of eigenvalues of different order. Hence, the curious behavior of the decay rates associated with the Orr-Sommerfeld hydrodynamic operator, as illustrated in Fig.~2 of paper I~\cite{miORR} or in Fig.~2 of Gustavsson and Hultgren~\cite{GustavssonHultgren}, is absent in the case of the decay rates associated with the Squire hydrodynamic operator. The general nature of the eigenvalue map for the Squire operator is the same as depicted in Fig.~\ref{F1} here, independent of the Reynolds number.

In view of the structure~\eqref{E72} of the eigenfunctions and of the boundary conditions $R_N(q_\parallel,\pm1)=0$, the normalization constants~\eqref{E60} can be evaluated exactly by using both the known integrals of products of the Airy functions~\cite{Albright} and the Wronskian of Airy functions. This procedure yields:
\begin{equation}\label{E19B}
B_N(q_\parallel)=\int_{-1}^{1}dz~[R_N(q_\parallel,z)]^2=\frac{-\mathrm{i}}{\pi^2(q_x\mathrm{Re})^{1/3}} \left\{ 1-\frac{\mathrm{Ai}\left[(q_x\mathrm{Re})^{1/3}(a_N-\mathrm{i})\right]} {\mathrm{Ai}\left[(q_x\mathrm{Re})^{1/3}(a_N+\mathrm{i})\right]}\right\},
\end{equation}
and a similar expression in terms of the Bi functions, see Eq.~\eqref{E74}. Again, as a consequence of the Squire symmetry, the normalization constant $B_N(q_\parallel)$ only depends on the product  $q_x\mathrm{Re}=q_\parallel\overline{\text{Re}}$.

\section{Nonequilibrium Vorticity Fluctuations\label{S04}}

Starting from the general theory presented in Sect.~\ref{S02B} and using the expressions for the hydrodynamic modes and decay rates derived in Sect.~\ref{S3} we can now evaluate the autocorrelation function $\langle\delta{w}^*_z(\omega,\mathbf{q}_\parallel,z)\cdot
\delta{w}_z(\omega^\prime,\mathbf{q}_\parallel^\prime,z^\prime)\rangle$ of the wall-normal vorticity fluctuations. By combining Eqs.~\eqref{EX33}, \eqref{E10}, and~\eqref{E13}, one readily obtains:
\begin{equation}\label{E19}
\langle\delta{w}^*_z(\omega,\mathbf{q}_\parallel,z)\cdot
\delta{w}_z(\omega^\prime,\mathbf{q}_\parallel^\prime,z^\prime)\rangle = W_{zz}(\omega,q_\parallel,z,z^\prime) ~(2\pi)^3~\delta(\omega-\omega^\prime)~\delta(\mathbf{q}_\parallel-\mathbf{q}_\parallel^\prime)
\end{equation}
with
\begin{equation}\label{E20}
W_{zz}(\omega,q_\parallel,z,z^\prime)=\sum_{N,M=0}^\infty \frac{\Xi^\text{(E)}_{NM} + \Xi^\text{(NE)}_{NM}(\omega)}{B_N^*~B_M[-\mathrm{i}\omega+\Gamma_N^*][\mathrm{i}\omega+\Gamma_M]}~R_N^*(z)~R_M(z^\prime).
\end{equation}
To simplify the notation we have suppressed on the RHS of Eq.~\eqref{E20} the explicit dependence of the quantities on the wave vector $\mathbf{q}_\parallel$. In principle, Eq.~\eqref{E20} enables us to investigate not only the intensity of the vorticity fluctuations, but also the time-dependent correlation function characterizing the dynamics these  fluctuations. However, just as in the case of the velocity fluctuations derived in paper~I~\cite{miORR}, we consider here only the intensity of the nonequilibrium vorticity fluctuations, which is given by the equal-time correlation function $\langle\delta\omega^*_z(\mathbf{q}_\parallel,z,t)\cdot\delta{w}_z(\mathbf{q}_\parallel^\prime,z^\prime,t)\rangle$. This equal-time correlation function is obtained by applying a double inverse Fourier transform, in frequencies $\omega$ and $\omega^\prime$, to Eq.~\eqref{E19} so that
\begin{equation}\label{E21}
\langle\delta\omega^*_z(\mathbf{q}_\parallel,z,t)\cdot\delta{w}_z(\mathbf{q}_\parallel^\prime,z^\prime,t)\rangle = W_{zz}(\mathbf{q}_\parallel,z,z^\prime)~(2\pi)^2~\delta(\mathbf{q}_\parallel-\mathbf{q}_\parallel^\prime),
\end{equation}
with
\begin{equation}\label{E22}
W_{zz}(\mathbf{q}_\parallel,z,z^\prime)=\frac{1}{2\pi} \int_{-\infty}^{\infty}d\omega~W_{zz}(\omega,\mathbf{q}_\parallel,z,z^\prime).
\end{equation}
Substituting Eq.~\eqref{E20} into Eq.~\eqref{E22}, and performing the integration over the frequency $\omega$ of the fluctuations, one readily obtains an expression for the amplitude of the equal-time correlation function. In view of the structure of Eq.~\eqref{E20},  we conclude that the resulting expression will contain two additive contributions, namely an equilibrium contribution (E) and a nonequilibrium contribution (NE):
\begin{equation}\label{E29B}
W_{zz}(\mathbf{q}_\parallel,z,z^\prime)= W_{zz}^\text{(E)}(\mathbf{q}_\parallel,z,z^\prime) + W_{zz}^\text{(NE)}(\mathbf{q}_\parallel,z,z^\prime).
\end{equation}

\subsection{Equilibrium contribution to the intensity of the fluctuations}

Since the mode-coupling coefficients $\Xi^\text{(E)}_{NM}$ do not depend on the frequency $\omega$, the integration in Eq.~\eqref{E22} can be readily performed for the equilibrium contribution. Combining the result with the expression~\eqref{E15} for the ``equilibrium" mode-coupling coefficients we arrive at:
\begin{equation}\label{E23}
\begin{split}
W_{zz}^\text{(E)}(\mathbf{q}_\parallel,z,z^\prime)&=\tilde{S}\text{Re}~q_\parallel^2 \sum_{N,M=0}^\infty \int_{-1}^{1}d\xi\frac{R_N^*(\xi)~R_M(\xi)}{B_N^*~B_M}~R_N^*(z)~R_M(z^\prime),\\
&=\tilde{S}\text{Re}~q_\parallel^2 \sum_{N=0}^\infty \int_{-1}^{1}d\xi\frac{R_N^*(\xi)}{B_N^*}~R_N^*(z)~\delta(\xi-z^\prime),\\
&=\tilde{S}\text{Re}~q_\parallel^2 ~\delta(z-z^\prime),
\end{split}
\end{equation}
where we have made use of the expansion of the delta function in terms of the hydrodynamic modes $R_N(\mathbf{q}_\parallel,z)$,
\begin{equation}\label{E24}
\delta(\xi-z)=\sum_{N=0}^\infty \frac{1}{B_N(q_\parallel)}~R_N(q_\parallel,\xi)~R_N(q_\parallel,z),
\end{equation}
which is obtained by using the orthogonality condition~\eqref{E60}. It is obvious that $\delta(\xi-z)$, as a function of $z$, satisfies the relevant boundary conditions. Expression~\eqref{E24} is valid for any value of the wave number $q_\parallel$ or the effective Reynolds number $\overline{\text{Re}}$. The important result is that the expression~\eqref{E23} for $W_{zz}^\text{(E)}(\mathbf{q}_\parallel,z,z^\prime)$ indeed reproduces the intensity of the equal-time autocorrelation function of the wall-normal vorticity fluctuations for a fluid in equilibrium. Note that the prefactor $\tilde{S}\text{Re}$ appears in Eq.~\eqref{E23} as a consequence of the adoption of dimensionless variables; when one reverts to physical variables the resulting prefactor is indeed independent of the shear rate.

Equation~\eqref{E23} justifies the use of superscripts (E) and (NE) in Eq.~\eqref{E13}. In addition, it shows that the nonequilibrium contribution to the intensity of vorticity fluctuations arises only from the coupling with the wall-normal velocity fluctuations in the stochastic Squire equation~\eqref{Eq09B}, and not from the self-coupling also present in the stochastic Squire equation~\eqref{Eq09B}. This result was already found previously~\cite{miCouette2} on the basis of a Galerkin approximation. We now see that this is exact, and not just a consequence of the simplicity of the approximation used previously~\cite{miCouette2}. However, the self-coupling in the Squire contribution does contribute to the time-dependent nonequilibrium correlation function which is not considered here.

The separation~\eqref{E29B} of the effects of thermal noise into an equilibrium and a nonequilibrium contribution is equivalent to what was found in the study of stochastic forcing, for instance, Eq.~(17) of Ref.~\cite{BamiehDahleh} or Eq.~(4.3) of Jovanovic and Bamieh~\cite{JovanovicBamieh}. The terms proportional to the cube of the Reynolds number in those expressions~\cite{BamiehDahleh,JovanovicBamieh} vanish when there is no flow in the system, but physically, the terms proportional to Re exists even when there is no flow. Indeed, as is the case in our Eq.~\eqref{E23}, one power of $\text{Re}$ in Refs.~\cite{BamiehDahleh,JovanovicBamieh} is due to dimensionality reasons. A forcing (stochastic or not) must have units of force and the dimensionless time is in unit of the (inverse) shear rate. An important feature of our calculation is that we recover the well-known expression~\eqref{E23} for the vorticity fluctuations in equilibrium~\cite{BoonYip,HansenMcDonald}.

\subsection{Nonequilibrium contribution to the intensity of the fluctuations}

Upon substitution of the (NE) part of Eq.~\eqref{E20} into Eq.~\eqref{E22}, one obtains the nonequilibrium contribution  $W_{zz}^\text{(NE)}(\mathbf{q}_\parallel,z,z^\prime)$ to the vorticity fluctuations. By making further use of Eq.~\eqref{E14} for the mode-coupling coefficients, we arrive at the explicit expression:
\begin{multline}\label{E26}
W_{zz}^\text{(NE)}(z,z^\prime)=\frac{q_y^2}{2\pi} \sum_{N,M=0}^\infty\hspace*{-6pt} \frac{R_N^*(z)~R_M(z^\prime)}{B_N^*~B_M}\iint_{-1}^1\hspace*{-6pt}d\xi d\xi^\prime R_N^*(\xi)~R_M(\xi^\prime)\\ \times \int_{-\infty}^{\infty} \frac{C_{zz}(\omega,\xi,\xi^\prime)~d\omega} {[-\mathrm{i}\omega+\Gamma_N^*][\mathrm{i}\omega+\Gamma_M]}.
\end{multline}
In principle one needs to substitute the autocorrelation function $C_{zz}(\omega,\xi,\xi^\prime)$ of the wall-normal velocity fluctuations from Paper~I~\cite{miORR} into Eq.~\eqref{E26}, and perform the various integrations and summations to obtain the function of interest. This procedure yields rather complicated expressions and only marginal analytical progress can be made, at the expense of very large and cumbersome expressions. Therefore,  unlike our previously solution obtained on the basis of a Galerkin approximation~\cite{miCouette2}, only a numerical computation of $W_{zz}^\text{(NE)}(z,z^\prime)$ is generally possible, and even this numerical procedure turns out to be rather long and difficult.

There is one important issue that should be mentioned. Because of the presence of the term $q_y^2$ as a prefactor in Eq.~\eqref{E26}, it turns out that the intensity of the nonequilibrium vorticity fluctuations has a maximum in the spanwise direction ($q_x=0$), while it is zero in the streamwise direction. This is opposite to the wave-vector dependence of the wall-normal velocity fluctuations discussed in Paper~I~\cite{miORR}. In that case the intensity of the fluctuations has a maximum in the streamwise direction ($q_y=0$) and is zero in the spanwise direction. Moreover, for the same Reynolds number the intensity of the vorticity fluctuations is substantially larger than the intensity of the wall-normal velocity fluctuations. Hence, an important conclusion that can be derived from Eq.~\eqref{E26} is that the most important effect of the flow on thermal fluctuations is the enhancement of wall-normal vorticity fluctuations with wave vector in the spanwise direction; or, equivalently, fluctuations that are constant in the streamwise direction. The same conclusion was obtained from our previous approximate Galerkin solution~\cite{miCouette,miCouette2}, see in particular Fig.~6 in Ref.~\cite{miJNNFM}. A similar conclusion is obtained both from direct numerical simulations of the full Navier-Stokes equations or from analytical studies of transient growth (amplification) of perturbations, see, for instance refs.~\cite{GaymeEtAl1,HwangCossu1}

Not only are the vorticity fluctuations in the spanwise direction the most dominant and interesting. In addition, further analytical progress is also possible for this case. This is the reason why previous investigators, who considered externally imposed forcing, have focused on vorticity response in the spanwise direction~\cite{FarrellIoannou,BamiehDahleh,EckhardtPandit}. Indeed, when $q_x =0$ the (Fourier transformed) Orr-Sommerfeld and Squire equations simplify notably. The eigenfunctions of the Squire operator can be simply expressed in terms of trigonometric functions, while the corresponding eigenvalues can be obtained analytically. The eigenfunctions of the Orr-Sommerfeld operator in this limit can be written as combinations of trigonometric and hyperbolic functions, while the eigenvalues can be obtained numerically by solving relatively simple algebraic equations, as first discussed by Dolph and Lewis~\cite{DolphLewis}. One important property is that in both cases the eigenfunctions possess a well-defined parity, and can be naturally classified into odd eigenfunctions and even eigenfunctions. We note that this is not true in the general case $q_x\neq0$. Hence, in the remainder of this paper we shall focus on the vorticity fluctuations in the spanwise direction, and their effect on the nonequilibrium energy amplification induced by the fluid flow.

\section{Nonequilibrium Energy Amplification\label{S05}}

We recall that in the derivation of the stochastic Orr-Sommerfeld and Squire equations we have used the incompressibility assumption $\boldsymbol{\nabla}\cdot\delta\mathbf{v}=0$~\cite{miCouette,miCouette2}. Hence, only two components of the velocity fluctuate independently. Most investigators on the subject have been interested in the so-called kinetic-energy amplification, that can be obtained from the sum of the equal-time autocorrelation functions $\langle\delta{v}^*_i(\mathbf{q}_\parallel,z,t)\cdot\delta{v}_i(\mathbf{q}_\parallel^\prime,z^\prime,t)\rangle$.
And indeed, as elucidated in more detail in our previous analysis on the basis of a Galerkin approximation~\cite{miCouette2}, the spatial spectrum of the kinetic-energy amplification is proportional to the vertical average
\begin{align}
\frac{1}{2}&\int_{-1}^{1}\int_{-1}^{1}\hspace*{-6pt}dz~dz^\prime~\sum_i
\langle\delta{v}^*_i(\mathbf{q}_\parallel,z,t)\cdot
\delta{v}_i(\mathbf{q}_\parallel^\prime,z^\prime,t)\rangle\label{Ex42}
\\
&=\frac{1}{2}\int_{-1}^{1}\int_{-1}^{1}\hspace*{-6pt}dz~dz^\prime \left\{\frac{1}{q_\parallel^2}
\langle\delta{w}^*_z(\mathbf{q}_\parallel,z,t)\cdot
\delta{w}_z(\mathbf{q}_\parallel^\prime,z^\prime,t)\rangle
+\langle\delta{v}^*_z(\mathbf{q}_\parallel,z,t)\cdot\delta{v}_z(\mathbf{q}_\parallel^\prime,z^\prime,t)\rangle\right\},\notag
\end{align}
where on the RHS of this equation we only need to consider the wall-normal velocity and vorticity fluctuations because of the divergence-free condition $\boldsymbol{\nabla}\cdot\delta\mathbf{v}=0$.  In addition, the fact that there is no cross-correlation between $\delta{w}_z$ and $\delta{v}_z$ has also been employed.

The contribution in Eq.~\eqref{Ex42} arising from the wall-normal velocity fluctuations has been studied extensively in Paper~I~\cite{miORR}. We focus here on the contribution from the vorticity fluctuations which, in view of Eqs.~\eqref{E21}-\eqref{E22}, will be proportional to the quantity
\begin{equation}\label{E28}
W_{zz}(\mathbf{q}_\parallel) = \frac{1}{2} \int_{-1}^{1}\int_{-1}^{1}\hspace*{-6pt}dz~dz^\prime~W_{zz}(\mathbf{q}_\parallel,z,z^\prime),
\end{equation}
which is a function of the horizontal wave vector $\mathbf{q}_\parallel$ of the fluctuations that we shall investigate in this section.

First of all, one notes that because of the structure of the mode-coupling coefficients, the spectrum $W_{zz}(\mathbf{q}_\parallel)$ can be expressed as the sum of an equilibrium and a nonequilibrium contribution which we prefer to write in the form
\begin{equation}\label{E29}
W_{zz}(\mathbf{q}_\parallel)=W_{zz}^\text{(E)}({q}_\parallel)\left[1+\Delta{W_{zz}^\text{(NE)}}(\mathbf{q}_\parallel)\right],
\end{equation}
where $\Delta{W_{zz}^\text{(NE)}}(\mathbf{q}_\parallel)$ represents the nonequilibrium energy enhancement. The equilibrium contribution $W_{zz}^\text{(E)}({q}_\parallel)$ in Eq.~\eqref{Ex42} is obtained by substituting Eq.~\eqref{E23} into Eq.~\eqref{E28}
\begin{equation}\label{E36}
W_{zz}^\text{(E)}({q}_\parallel)=\tilde{S}\text{Re}~q_\parallel^2,
\end{equation}
which is the same as in the absence of any flow.

The nonequilibrium enhancement $\Delta{W_{z}^\text{NE}}(\mathbf{q}_\parallel)$ in Eq.~\eqref{E29}, resulting from the wall-normal vorticity fluctuations, can be obtained by substituting Eq.~\eqref{E26} into Eq.~\eqref{E28}. As already mentioned before, in general, this procedure yields a complicated expression that can only be evaluated numerically.

\subsection{Enhancement of streamwise-constant fluctuations (\emph{i.e.}, with wave vector in the spanwise direction)}

A particular simple case is that of fluctuations constant in the streamwise direction with $\mathbf{q}$ in the spanwise direction, \emph{i.e.}, for which $q_x=0$ and $q_y=q_\parallel=q$. As mentioned above, under this condition the working equations simplify greatly, and a more compact analytical expression can be obtained for the nonequilibrium enhancement. Because of its obvious simplicity, this particular case has been analyzed in detail by some previous investigators~\cite{FarrellIoannou,BamiehDahleh,JovanovicBamieh} but for externally imposed stochastic forcing, not thermal noise. Hence, their results differ from the ones obtained here. For this reason, we present our explicit results for the enhancement of vorticity fluctuations induced by thermal noise with the (horizontal) wave vector in the spanwise direction, namely:
\begin{multline}
\left.\Delta{W_{zz}^\text{NE}}(q)\right|_{q_x=0} =\frac{\text{Re}^2}{8q^3}\left[\frac{9-\tanh^2q}{2q}-\frac{\tanh{q}}{\cosh^2q}-\frac{9\tanh{q}}{2q^2}\right]\\ + \text{Re}^2 \sum_{N=0}^\infty\frac{(2a_N^2+3q^2)\tanh\sqrt{a_N^2+2q^2}+\dfrac{q^2\sqrt{a_N^2+2q^2}}{\cosh^2\sqrt{a_N^2+2q^2}}}{(a_N^2+q^2)^2(a_N^2+2q^2)^2\left[\dfrac{q\tanh{q}}{\sqrt{a_N^2+2q^2}}- \tanh\sqrt{a_N^2+2q^2} \right]},\label{E37}
\end{multline}
with
\begin{equation}
a_N=\frac{\pi}{2}(2N+1),
\end{equation}
so that $(a_N^2+q^2)/\text{Re}$ are the eigenvalues of the Squire operator in the spanwise direction with  corresponding eigenfunctions of even parity. As already mentioned, in this particular case (spanwise direction) the hydrodynamic modes (eigenfunctions) have a well-defined parity. Because of the $z$-integrations in Eq.~\eqref{E14B} for the mode-coupling coefficients, modes with different parity do not couple. In addition, because of the integrations in Eq.~\eqref{E28}, only the even eigenfunctions or modes do finally contribute to the nonequilibrium enhancement. To obtain Eq.~\eqref{E37} we have closely followed a procedure used by Bamieh and Dahleh~\cite{BamiehDahleh}; in particular we used the auxiliary function $g(z)$ introduced in their \emph{Lemma 4}. However, we obtain a different result because the thermal noise, considered here, has a special spatial spectrum given by the fluctuation-dissipation theorem~\eqref{FDT1}, that is different from the spectrum of the externally imposed stochastic forcing considered by Bamieh and Dahleh~\cite{BamiehDahleh}. One power of the Reynolds number appears in the energy amplification because of dimensionality reasons. Physically, the flow-induced amplification of thermal noise is proportional to the square of the Reynolds number (shear rate), not to the cube of the Reynolds number as stated by Bamieh and Dahleh~\cite{BamiehDahleh} or Jovanovic and Bamieh~\cite{JovanovicBamieh}.
\begin{figure}
\begin{center}
\resizebox{0.60\textwidth}{!}{\includegraphics{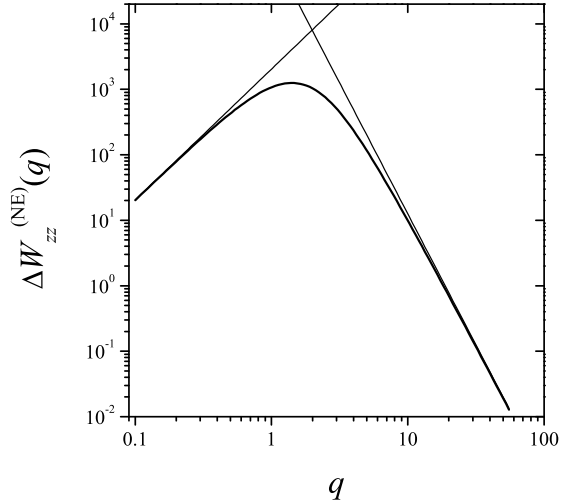}}
\end{center}
\caption{Nonequilibrium enhancement $\Delta{W_{z}^\text{NE}}(q)$ of wall-normal vorticity fluctuations in the spanwise direction ($q_x=0$) from Eq.~\eqref{E37}. The two straight lines represent the asymptotic behaviors at small and large wave numbers, Eqs.~\eqref{E39} and~\eqref{E40}, respectively. Data are for $\text{Re}=500$.}\label{F2}
\end{figure}

In Fig.~\ref{F2} we show a plot of the nonequilibrium enhancement $\Delta{W_{zz}^\text{NE}}(q)$ of the wall-normal vorticity fluctuations with wave vector in the spanwise direction ($q_x=0$, or streamwise constant) as given by Eq.~\eqref{E37}. The data in Fig.~\ref{F2} are for $\text{Re}=500$, but we note that the Reynolds number appears in Eq.~\eqref{E37} only as as a prefactor, so that the ratio $\Delta{W_{zz}^\text{NE}}(q)/\text{Re}^2$ does not depend on the Reynolds number. We conclude from Fig.~\ref{F2} that the spanwise energy amplification can be very well visualized as a simple crossover between the two asymptotic behaviors, at large and small $q$, that can be easily obtained from Eq.~\eqref{E37}, namely
\begin{align}
\left.\Delta{W_{zz}^\text{NE}}(q)\right|_{q_x=0}\xrightarrow{q\to{0}}&~\text{Re}^2q^2\left[\frac{34}{315}-\frac{256}{\pi^7} \sum_{N=0}^\infty \frac{2+\cosh{(2N+1)\pi}}{(2N+1)^7 \sinh{(2N+1)\pi}} \right]\notag\\&~\simeq 8.14\times10^{-3}~\text{Re}^2q^2,\label{E39}\\
\left.\Delta{W_{zz}^\text{NE}}(q)\right|_{q_x=0}\xrightarrow{q\to{\infty}}&~\frac{\text{Re}^2}{2q^4}.\label{E40}
\end{align}
We recover in Eq.~\eqref{E40} the $q^{-4}$ behavior that is typical of nonequilibrium fluctuations~\cite{BOOK} at large wave numbers corresponding to wavelengths smaller than the spacing between the walls. For such wavelengths the fluctuations are not affected by the boundary conditions and we recover, as an asymptotic limit for large $q$, results obtained by previous investigators for nonequilibrium fluctuations in fluids under shear~\cite{TremblayEtAl,LD85,LDD89,LD02,WS03}. On the other hand, the vanishing of the intensity of fluctuations with small $q$, as implied by Eq.~\eqref{E39}, is to be expected in the sense that the walls (boundary conditions) effectively suppress fluctuations of very long wavelength, comparable with the separation distance between walls. From Fig.~\ref{F2} we note that the flow at $\text{Re}=500$ causes an enhancement of the thermal energy about 1000 times larger that the thermal energy that would be expected from a local equilibrium assumption. Such a profound enhancement of the fluctuations is a general phenomenon in fluids in nonequilibrium states~\cite{BOOK}.

\begin{figure}
\begin{center}
\resizebox{0.60\textwidth}{!}{\includegraphics{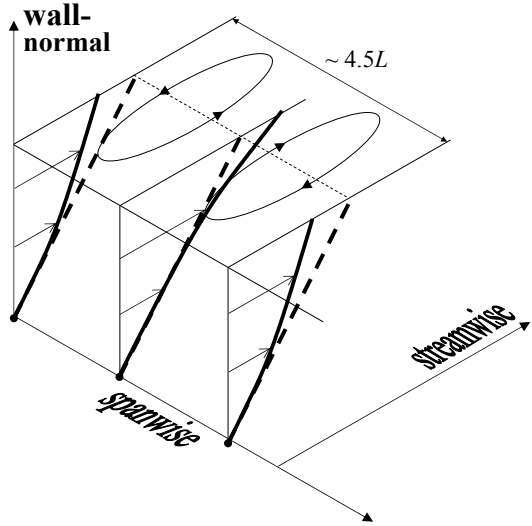}}
\end{center}
\caption{When we add to the mean Couette flow a set of wall-normal vortices with spanwise modulation of about $4.5L$ and very elongated in the streamwise direction, narrow regions where the streamwise velocity is larger or smaller than the average develop, customarily referred to as streaks. For clarity, we draw these vortices only at a reference height, although in general they will present some vertical (wall-normal) extension.}\label{F2B}
\end{figure}

Figure~\ref{F2} shows that the main effect of the flow on the thermal fluctuations is to select and maximally amplify the wall-normal vorticity fluctuations with a particular value of the wave vector $\mathbf{q}_\text{m}$. As already discussed, the wave vector maximally enhanced is in the spanwise direction, and from Eq.~\eqref{E37} we find numerically its magnitude to be $q_\text{m}\simeq1.4103$, which is the location of the maximum in Fig.~\ref{F2}. Therefore, in real space, the thermal noise amplified by the flow will manifest itself mainly as a set of vortices of size $\simeq(2\pi/1.41)L\simeq4.5L$ distributed in the spanwise direction and that are constant (extremely elongated) in the streamwise direction\footnote{This is by approximating the complicated spatial spectrum of the thermal velocity-fluctuations, as shown for instance in Fig.~6 of Ref.~\cite{miJNNFM}, by just two delta functions located at the symmetric maxima (of the fluctuating wall-normal vorticity) at $q_x=0$, $q_y=\pm1.4103$.}. The intensity of these fluctuating vortices will be proportional to $k_\text{B}T$ and to the square of the Reynolds number, see Eq.~\eqref{E37}. But we note that close to the center of the layer, where the base flow velocity is zero, the velocity fluctuations might be of the same order as the mean (base) velocity.  These fluctuating vortices deform the base flow, developing a series of streaks, \emph{i.e.}, narrow regions where the streamwise velocity is larger or smaller than the average, as indicated in Fig.~\ref{F2B} where we have tried to illustrate schematically this physical situation. These streaks will be typically separated by a distance of about 4.5 times the half gap between the walls, as also shown in Fig.~\ref{F2B}. This optimal spanwise wave number is independent of the Reynolds number.

At this point it is interesting to note that several authors have identified the appearance of a set of fluctuating streaks in sheared flows as a precursor of the instability~\cite{Waleffe1,HofEtAl}. More quantitatively, extensive numerical simulations of the unstabilization of plane Couette flow~\cite{KomminahoEtAl,DuguetEtAl,GaymeEtAl1,HwangCossu1} have revealed large-scale coherent streaks with typical spanwise wavelength of $\approx3L-4L$. For instance, most recently Gayme et al.~\cite{GaymeEtAl1} performed an extensive analysis of direct numerical simulation data by Tsukahara et al.~\cite{TsukaharaEtAl}, obtaining an optimal spanwise wavelength of 1.8 times the distance between the plates, or 3.6 times the half distance, to be compared with our result of $4.5L$. Of course, the scope of our present work is restricted by the use of linear equations, so that its relevance to shear flow instability has to be considered, at most, as tentative. It is well-known and widely accepted that a complete understanding of unstabilization of shear flows requires a fully nonlinear theory~\cite{Waleffe1,HofEtAl}. In any case, the identification of the modes that are maximally enhanced in a linear theory may provide useful insights for developing simplified nonlinear theories, like the single mode model recently discussed by Gayme et al.~\cite{GaymeEtAl1}. Furthermore, it is intriguing to know that a linear theory predicts that thermal (natural) noise develops into ``streaks", of intensity $\propto k_\text{T}\text{Re}^2$, separated by a distance about $4.5L$, in agreement with numerical simulations of the nonlinear problem~\cite{HwangCossu1}.

To conclude this section, we mention that the algebraic $\propto{q}^{-4/3}$ dependence found for the enhancement of bulk fluctuations vanishes in the spanwise direction~\cite{miCouette2}. That is the reason why the ``shoulder" at intermediate $q$ shown in the bottom panel of Fig.~2 of Ref.~\cite{miCouette2} does not appear in our current results. Bulk nonequilibrium fluctuations in sheared fluids have been first investigated by Tremblay et al.~\cite{TremblayEtAl}, who described the $q^{-4}$ behavior. Subsequently, Dufty and Lutsko~\cite{LD85,LD02} included the algebraic wave-number dependence that, in general, appears at shorter wavelengths. Although these earlier papers~\cite{TremblayEtAl,LD85,LD02} refer specifically to wall-normal velocity fluctuations, we have found elsewhere~\cite{miCouette2} that these features are also present for vorticity fluctuations in bulk.

\subsection{Comparison with Galerkin approximation}

\begin{figure}
\begin{center}
\resizebox{0.60\textwidth}{!}{\includegraphics{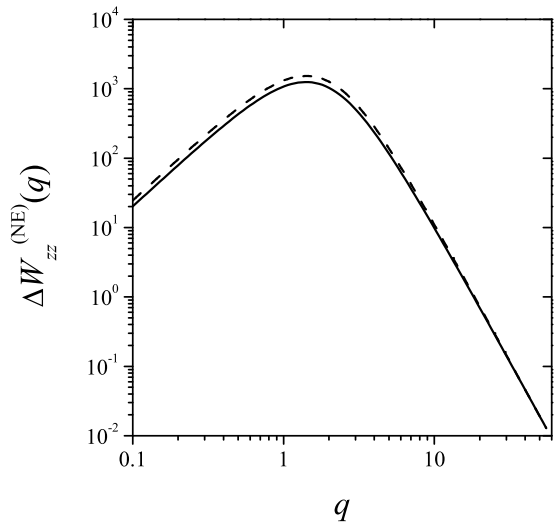}}
\end{center}
\caption{Comparison of the exact nonequilibrium enhancement $\Delta{W_{z}^\text{NE}}(q)$ of wall-normal vorticity fluctuations in the spanwise direction ($q_x=0$) from Eq.~\eqref{E37} (solid curve), with a Galerkin approximation developed in previous publications~\cite{miCouette2,miJNNFM}, see Eq.~\eqref{E41} (dashed curve). Data are for $\text{Re}=500$. The agreement is good.}\label{F3}
\end{figure}
In our previous publications~\cite{miCouette2,miJNNFM} the same problem considered here was investigated on the basis of a simple Galerkin approximation. In these previous publications we did not present explicit analytical expressions for the intensity of the vorticity fluctuations (on the basis of a Galerkin approximation), but only displayed the results graphically. However, as in the case of the exact solution, also in case of the Galerkin approximation the solution simplifies greatly for vorticity fluctuations in the spanwise direction. Specifically, for the fluctuations in the spanwise direction the expression in terms of the Galerkin approximation in Refs.~\cite{miCouette2,miJNNFM} reduces to
\begin{equation}\label{E41}
W_{zz}(q) = \tilde{S}\frac{5}{6} \text{Re}q^2\left[1+\frac{27\text{Re}^2q^2}{7(2q^2+5)(4q^4+23q^2+78)} \right].
\end{equation}
Equation~\eqref{E41} contains an equilibrium contribution that is about 17\% percent lower than the exact result, Eq.~\eqref{E36}. This is a shortcoming of the Galerkin-approximation scheme, as discussed elsewhere~\cite{miCouette2}. But our purpose here is to compare the nonequilibrium enhancement, given by the second term inside the square brackets in Eq.~\eqref{E41}, with the exact result given by Eq.~\eqref{E37}. For this purpose we show in Fig~\ref{F3} the exact $\Delta{W_{z}^\text{NE}}(q)$ in the spanwise direction from  Eq.~\eqref{E37} as a solid curve, together with the Galerkin approximation~\eqref{E41} of this quantity as a dashed curve. The agreement is rather good qualitatively. Quantitatively, the Galerkin approximation~\eqref{E41} for large $q$ underpredicts the exact asymptotic limit~\eqref{E40} by 4\% (indistinguishable on the scale of Fig.~\ref{F3}), whereas in the small $q$ limit Galerkin approximation overpredicts the exact asymptotic limit~\eqref{E41} by about 60\%. The prediction, on the basis of the Galerkin approximation~\eqref{E41}, of the wave number of maximum enhancement $q_\text{m}$ is quite good, differing by about 1\% from the true value.

Although we have here compared explicitly the exact solution with the Galerkin approximation only in the spanwise direction, from the good results obtained we may infer that the Galerkin approximation developed in Ref.~\cite{miCouette2} will also yield a good representation of the nonequilibrium enhancement of the wall-normal vorticity fluctuations in any direction. This expectation has been confirmed by some preliminary calculations for arbitrary $\mathbf{q}_\parallel$  on the basis of Eq.~\eqref{E26}.

\section{Concluding remarks\label{S07}}

We have shown that even in the absence of any external perturbations, already the thermally excited fluctuations, which are always present, cause a substantial energy amplification in laminar fluid flow with the main contribution to the energy amplification arising from wall-normal vorticity fluctuations as a result of a coupling of these wall-normal vorticity fluctuations with the wall-normal velocity fluctuations. On the other hand, in computational fluid dynamics, to destabilize shear flows and to investigate the transition to turbulence, some externally imposed random initial conditions are customarily introduced~\cite{DuguetEtAl}. We have demonstrated that such externally imposed initial conditions are not needed physically. Indeed, thermal noise causes already the base flow to spontaneously develop streaks, which are currently expected to be the first step in the transition to turbulence, even in a linear approximation. We have evaluated the typical spanwise distance between these thermally excited streaks, finding excellent agreement with what is observed in simulations. Of course, the problem of the nonlinear evolution of these streaks needs to be further investigated. For this purpose direct numerical simulations of fluctuating hydrodynamics may be of interest.

In the last few decades there has been considerable interest in the fluid dynamics community in these and similar problems, like optimal disturbances, transient growth of perturbations, and energy amplification. At the same time, and somewhat independently, in the statistical physics community there has been an interest in the nonequilibrium enhancement of thermal fluctuations. The purpose of this paper, as well of the previous paper~\cite{miORR}, has been to make a connection between the approaches used in statistical physics and fluid dynamics.

\begin{acknowledgements}

The authors acknowledge stimulating discussions with Andreas Acrivos, Bruno Eckhardt and Mihailo Jovanovic. The research was supported by the Spanish Ministry of Science and Innovation through research grant FIS2008-03801.

\end{acknowledgements}


\end{document}